\documentclass[conference]{IEEEtran}

\IEEEoverridecommandlockouts

\usepackage{cite}
\usepackage{amsmath,amssymb,amsfonts}
\usepackage{algorithmic}
\usepackage{graphicx}
\usepackage[table,xcdraw]{xcolor}
\usepackage{textcomp}
\usepackage{xcolor}
\def\BibTeX{{\rm B\kern-.05em{\sc i\kern-.025em b}\kern-.08em
    T\kern-.1667em\lower.7ex\hbox{E}\kern-.125emX}}
\usepackage{enumitem}

\usepackage[hyphens]{url}

\usepackage{xcolor}

\newcommand{\linebreakand}{%
  \end{@IEEEauthorhalign}
  \hfill\mbox{}\par
  \mbox{}\hfill\begin{@IEEEauthorhalign}
}

\title{A Blueprint for Collaborative Cybersecurity Operations Centres with Capacity for Shared Situational Awareness, Coordinated Response, and Joint Preparedness}

\begin{document}

\IEEEoverridecommandlockouts

\author{\IEEEauthorblockN{Konstantinos Fysarakis\IEEEauthorrefmark{1},
Vasileios Mavroeidis\IEEEauthorrefmark{2}, 
Manos Athanatos\IEEEauthorrefmark{3}, 
George Spanoudakis\IEEEauthorrefmark{1} and
Sotiris Ioannidis\IEEEauthorrefmark{3}}
\IEEEauthorblockA{\IEEEauthorrefmark{1}Sphynx Analytics Limited, Nicosia, Cyprus\\ Email: \{fysarakis, spanoudakis\}@sphynx.ch}
\IEEEauthorblockA{\IEEEauthorrefmark{2}University of Oslo, Oslo, Norway\\ Email: vasileim@ifi.uio.no}
\IEEEauthorblockA{\IEEEauthorrefmark{3}Technical University of Crete, School of Electrical \& Computer Engineering, Chania, Greece\\
Email: mathanatos@tsi.gr, sotiris@ece.tuc.gr}}

\IEEEpubid{%
\makebox[\columnwidth][l]{%
\raisebox{-14pt}[0pt][0pt]{%
\parbox[b]{\columnwidth}{%
\fontsize{5.5}{6.2}\selectfont
\raggedright
\textcopyright~2022 IEEE.
Personal use of this material is permitted.
Permission from IEEE must be obtained for all other uses, in any current or future media, including reprinting/republishing this material for advertising or promotional purposes, creating new collective works, for resale or redistribution to servers or lists, or reuse of any copyrighted component of this work in other works.
Published in the 2022 IEEE International Conference on Big Data (Big Data), pp.~2601--2609, 2022.
DOI:
\pdfstartlink
attr{/Border[0 0 0]}
user{/Subtype/Link/A<</S/URI/URI(https://doi.org/10.1109/BigData55660.2022.10020736)>>}
10.1109/BigData55660.2022.10020736
\pdfendlink
}%
}%
}%
\hspace{\columnsep}%
\makebox[\columnwidth]{}%
}

\maketitle  
\IEEEpubidadjcol

\begin{abstract}
With digital technologies now being part of the fabric of our societies, identifying and managing cybersecurity threats becomes imperative. Within the European Union, several initiatives are underway, aiming to motivate, regulate and eventually orchestrate the establishment of capacity and enhancement of situational awareness, incident response, and preparedness capabilities, with an expected emphasis on operators of essential services and state actors entrusted with cybersecurity. In this context, the institution of cooperation and information exchange channels to allow for coordinated cross-border responses to large-scale incidents is particularly prioritised. Motivated by the above, this work presents a conceptual blueprint in support of architecting and establishing interoperable Cyber Security Operations Centres that combine capacity for situational awareness, incident response, and preparedness, also benefiting from the interplay between them, ultimately enhancing national cybersecurity capabilities, cross-border collaboration, and national supervision of their critical sectors, in line with current and upcoming regulatory requirements and the ever-increasing need for national and international cooperation.
\end{abstract}


\section{Introduction}
%

As digital technologies, applications, and services become increasingly pervasive in everyday lives and economies, cybersecurity incidents are also getting more frequent and diversified. In 2020 alone, 3,932 publicly reported data breaches led to the exposure of 37 billion records \cite{incidents}. The European Union Agency for Cybersecurity (ENISA), in their 2020 \cite{ENISA1} and 2021 \cite{ENISA2021} ENISA Threat Landscape (ETL) reports, emphasised that in the current decade, cybersecurity risks will become harder to assess and interpret due to the growing complexity of the threat landscape, adversarial ecosystem, and expansion of the attack surface. In addition, they highlighted an increase in sophisticated and targeted ransomware in the public sector, critical infrastructures, and other key industries, and the increased prevalence of hybrid threats, combining both the cyber and physical dimensions. The 2022 ETL \cite{ENISA2022} further highlights the impact geopolitical tensions have had in this regard (e.g., the Russian-Ukraine conflict) and a common trend for state-sponsored threat actors and cyber criminals alike to increasingly focus on supply chain compromises and Operational Technology (OT) assets. The COVID-19 pandemic has also exacerbated the risk exposure, being a prolific period for malicious actors targeting sensitive areas (e.g., healthcare organisations) \cite{ENISA2}. 

Collaborative defence, technologically enabled by interoperable cyber defence ecosystems that facilitate the free exchange of information, insights, analytics, and response across tools and teams, can be an effective approach to identifying, assessing, and managing these risks and countering the associated threats. For instance, Security Operations Centres (SOCs) and operational teams like Computer Security Incident Response Teams (CSIRTs) should be able to communicate and competently cooperate, have a shared threat situational awareness, and jointly respond to cyber threats promptly and effectively \cite{ENISA3}. There is also a consensus that to be successful in this mission, defenders need to invest in three foundational pillars, namely technology (e.g., network and endpoint monitoring tools), processes (e.g., security policies and incident response plans with clearly defined roles, assigned responsibilities, activities to perform, and synergies between teams), and people (e.g., training analysts and responders) \cite{SANS}. However, while there is a plethora of resources and references on the aforementioned individual topics, there is a lack of efforts that comprehensively present and analyse their interplay \cite{MITRE}.

Motivated by the aforementioned areas of concern, defenders' pressing need to stay threat informed and prepared to respond to this ever-evolving threat landscape, and considering the European Union (EU) dimension, which concentrates on strengthening the Union's capacity and ability to perform joint cybersecurity operations (comprehensively discussed in the following sections), this work presents and describes a baseline blueprint to aid defenders in understanding how to architect and establish Cyber Security Operations Centres (CSOCs) for cross-border, cross-organisational, and cross-functional cooperation, collaboration, and coordination, combining capacity for threat situational awareness and sharing, incident handling, and preparedness. Finally, this work pinpoints \textit{standardisation's} central role in achieving this mandated level of interoperability across defence environments, tools, processes, and people.

The rest of the paper is structured as follows: Section \ref{section:background} details the CSOC background in terms of terminology and reference resources and provides an overview of key motivators behind this work, including the relevant EU cybersecurity strengthening strategies; Section \ref{section:blueprint} presents and discusses the capacity-focused blueprint for architecting collaborative CSOCs proposed herein; and Section \ref{section:conclusions} provides the concluding remarks, including pointers to next steps and the validation of the proposed blueprint.

\section{Background} \label{section:background}
\subsection{Terminology}

The term CSIRT was established in the 1990s to refer to a team assigned to handle computer security incidents. Since then, we have seen a surge in the use of the term to describe teams involved in a broader set of operational roles and services in addition to incident handling (see, for example, Figure \ref{fig:FIRST}). Similarly, when we refer to a SOC, the community used to extrapolate its function to the availability of analysts whose responsibilities fall within security monitoring and detection, event analysis, triaging and escalation. In other cases, a SOC can be tiered and offer more services, including incident handling, intelligence gathering and intrusion analysis, threat hunting and more. Consequently, such terms have become ambiguous, all-encompassing, and often used interchangeably. 
In fact, identifying the services offered by a team judging by the name alone is arguably not possible since no strict classification is followed. Further, the plethora of terms that describe relevant teams within organisations, such as Computer Emergency Response Team (CERT), Computer Incident Response Team (CIRT), Security Incident Response Team (SIRT), Cybersecurity Operations Center (CSOC), and Computer Incident Response Center (CIRC), add to this challenge \cite{MITRE}.

Aspiring to create a capacity-focused blueprint to support the establishment of an EU-wide infrastructure for collaborative cybersecurity operations (as per the requirements described in Section \ref{sec:EU} ), in this work, we adhere to the term CSOC for uniformity, and it is used to refer to an ecosystem comprising people, along with specific capacity (including infrastructure and technologies), and tailored processes to support:

\begin{enumerate}[label=(\roman*)]
    \item core and optional services expected of private and public CSIRTs and SOCs (a CSOC can comprise different specialised teams with clearly defined roles, duties, and modes of operation), and
    \item collaboration and coordination expected from public entities entrusted with cybersecurity in the EU (e.g., national CSIRTs and sectorial CERTs of Member States).
\end{enumerate}

\subsection{Reference Resources}
There is a plethora of resources focused on or related to CSIRTs and SOCs and the different services they typically offer. However, the majority of the identified material is either outdated or limited in scope, concentrating on specific tools or operational roles, services and processes and rarely attempting to provide a more comprehensive approach to all the above and their interplay.

\begin{figure*}[tbp]
\centering
\includegraphics[width=0.7\hsize]
     {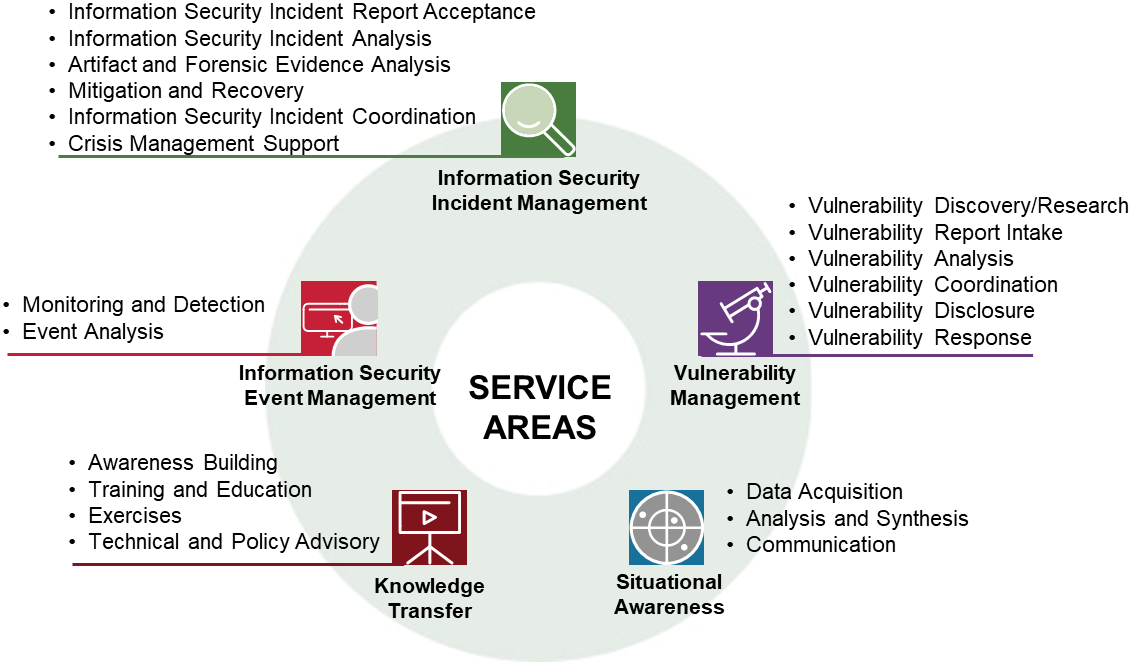}
\caption{Incident Response Teams' Service Areas and Services \cite{FIRST}.}
\label{fig:FIRST}
\end{figure*}

Nonetheless, for this work of importance are efforts specifying the potential services of CSIRTs and SOCs, as they facilitate the establishment or improvement of relevant operations. Further, they can also provide a valuable reference for terms and definitions that can be consistently used across the community. Such a prominent, widely recognised reference is the CSIRT Services Framework \cite{FIRST}, developed and maintained by the Forum of Incident Response and Security Teams (FIRST) and supported by ENISA, the International Telecommunications Union (ITU), and other organisations. The CSIRT Services Framework, at Version 2.1 since November 2019, is a high-level document describing a set of service areas and specific services under each area that CSIRTs may offer (see Figure \ref{fig:FIRST}). Teams can choose which are relevant to their mandate and incorporate them into their services portfolio.

Moreover, particularly important for the purpose of this work are guidelines on how to structure, set up, and operate CSIRTs, SOCs, and related teams. Noteworthy examples are the guidelines, and good practice guides on establishing and operating a CSIRT provided by ENISA and the U.S. National Institute of Standards and Technology (NIST), \textit{"How to setup CSIRT and SOC - Good Practice Guide"} \cite{ENISA3} and SP 800-61 Rev. 2 \textit{"Computer Security Incident Handling Guide"} \cite{NIST}, respectively.

A notable resource in terms of comprehensiveness and the holistic approach and perspective it provides regarding cybersecurity operations and the establishment of CSOCs is published by MITRE \cite{MITRE}. The requirements of establishing a state-of-the-art ("world class", per the book's wording) CSOC are presented from a strategic perspective, detailing ten strategies and how these crosscut technologies, processes, and people. In fact, the first of the strategies presented in the document is also at the heart of the concept presented herein: \textit{"Consolidate functions of incident monitoring, detection, response, coordination, and computer network defence tool engineering, operation, and maintenance under one organization: the CSOC"}.

The above references can be augmented by additional resources on incident management processes \cite{ENISA6} and other services, such as digital forensics \cite{ENISA7}. There is also significant work on maturity models \cite{SIM3}\cite{ENISA-MaturityModel}, and their counterpart evaluation methodologies \cite{ENISA-MaturityEval} that CSIRTs and SOCs can leverage to measure their capability and maturity.

Furthermore, numerous identified resources are focused on specific technologies and tools that CSIRTs and SOCs leverage to carry out their mission, such as for malware analysis \cite{Malware}, vulnerability assessment and penetration testing \cite{Vulnerability}, and network \cite{Network} and endpoint monitoring \cite{Host}. However, in the context of this work, these are very narrow in scope.

To sum up, other than some notable exceptions, most identified resources cover only a subset of the services that a CSOC could support and/or cover specific technologies and tools, without considering people and processes. In contrast, other resources cover people and processes without considering technologies and tools.

\subsection{The EU Perspective and Capacity Requirements} \label{sec:EU}

ENISA \cite{ENISA1} emphasises that dealing with systemic and complex risks is a top cybersecurity challenge since the increased motivation (often combining financial, nation-state interests, nationalistic and political motives) and sophistication of threat actors (often state-sponsored) carrying out targeted and persistent attacks, combined with the ever-increasing inter-connectivity of various systems and networks, make cyber risks harder to assess and mitigate. In this regard, ENISA's policy \cite{ENISA4} conclusions and recommendations highlight the importance of using and exchanging Cyber Threat Intelligence (CTI) for cybersecurity preparedness and for driving strategic and political decisions that can effectively tackle threats endangering the well-being of the EU. It is also emphasised that cooperation, collaboration, and coordination of EU-wide CTI activities are essential for informing and driving emergency decisions needed in crisis management. Moreover, having access to a cyber range and other structured training means are referred to as essential enablers for building, enhancing, and maturing the skills of security teams by simulating attacks and testing multiple defence strategies.

From a legislation and policy perspective, the EU, to establish a common high level of cybersecurity and foster coordination and cooperation across Member States, their Operators of Essential Services (OES), and Digital Service Providers (DSP), introduced the Directive on Security of Network and Information Systems (NIS Directive, 2016/1148) \cite{NIS}. In response to the directive, many competent authorities and teams were established across the EU and were entrusted with their national and critical infrastructure security. The NIS Directive also instituted the EU CSIRTs Network \cite{CSIRTSnet} to develop confidence and trust between the Member States and promote operational cooperation. The CSIRTs Network comprises EU Member States’ appointed CSIRTs, CERT-EU \cite{CERTEU}, and is supported by ENISA.

In addition, the EU Cybersecurity Strategy \cite{EUStrat} adopted in December 2020 further reinforced the focus on collective resilience against cyber threats, including strengthening collective capabilities to respond to major cyberattacks. In fact, national cybersecurity strategies \cite{NCSS} of Member States align with the EU Cybersecurity Strategy, covering international cooperation, capacity and capability for incident response, reporting mechanisms, trusted information-sharing, organisation of cybersecurity exercises, and also the strengthening of training and educational programmes. In line with this strategic direction, several initiatives are underway within the EU to address cybersecurity challenges in services critical to the economy and the broader society, such as the European Cyber Resilience Act, NIS2 \cite{NIS2} (an update of NIS Directive which further reinforces and elaborates upon the cybersecurity requirements imposed upon OES, while also incorporating more entities into the OES category), and the establishment of a Joint Cyber Unit \cite{JCU}.

The above are very much aligned with the European Commission Recommendation on Coordinated Response to Large Scale Cybersecurity Incidents and Crises \cite{blueprint}, referred to henceforth as Cyber Blueprint. The Cyber Blueprint describes high-level standard operating "procedures" pertinent to responding to large-scale cybersecurity incidents and crises. It specifies required capacity and capabilities, including shared situational awareness, preparedness and coordinated response, and pinpoints crisis management and cooperation mechanisms along with the key involved EU cybersecurity actors like ENISA, the CSIRTs Network, CERT-EU, and Europol. Of particular interest in this regard is the Cyber Crisis Liaison Organisation Network (CyCLONe \cite{CyCLONe}) that aims to contribute to the implementation of Cyber Blueprint and complements the existing EU cybersecurity structures by linking cooperation at technical, operational, and political-strategic levels. In fact, NIS2 will formally establish the EU-CyCLONe network, which will support the coordination and management of large-scale incidents (in line with NIS2's third objective \cite{NIS2} to improve the level of joint situational awareness at the EU level and the collective capability to prepare and respond).

The ENISA survey "Study on CSIRT landscape and Incident Response (IR) capabilities in Europe 2025", covering 444 CSIRTs, revealed that the legal and regulatory drivers marshalled by the implementation of the NIS Directive have a positive effect on the adoption of a holistic approach towards IR. Moreover, these drivers facilitate the alignment of Member States' national capabilities and have a positive effect at an international level \cite{ENISA5}.

By reconciling the above information, this work aimed to highlight the EU's cybersecurity focus towards establishing, enhancing, and harmonising mechanisms for cross-border preparedness, shared threat situational awareness, and coordinated response. The capacity-focused blueprint for collaborative CSOCs presented and discussed in the next section concentrates on the above needs and priorities.

\section{A Capacity-Focused Blueprint for Cross-Border Collaborative Cybersecurity Operations Centres}
\label{section:blueprint}

Hereafter, we present and discuss a capacity-focused blueprint that provides baseline guidelines for designing collaborative CSOCs. In the context of this work, we focus on the following capacity pillars and their interplay, also considering their intersection with technology, processes, and people. In particular:

\begin{enumerate}[label=(\roman*)]
\item \textit{situational awareness} that is accurate and up-to-date, supported by relevant, actionable, and timely cyber threat intelligence;
\item \textit{incident response} allowing for the timely execution of effective response strategies at the technical, operational and political-strategic levels \cite{blueprint}, covering different types of threats and scales (from isolated to large-scale); and 
\item \textit{preparedness} by leveraging a multi-faceted training approach to enhance defenders' capabilities, including educational material and realistic applied and adaptive training driven by (i) and (ii).
\end{enumerate}

\begin{figure}[!h]
\begin{center}
	\includegraphics[width=0.9\columnwidth]{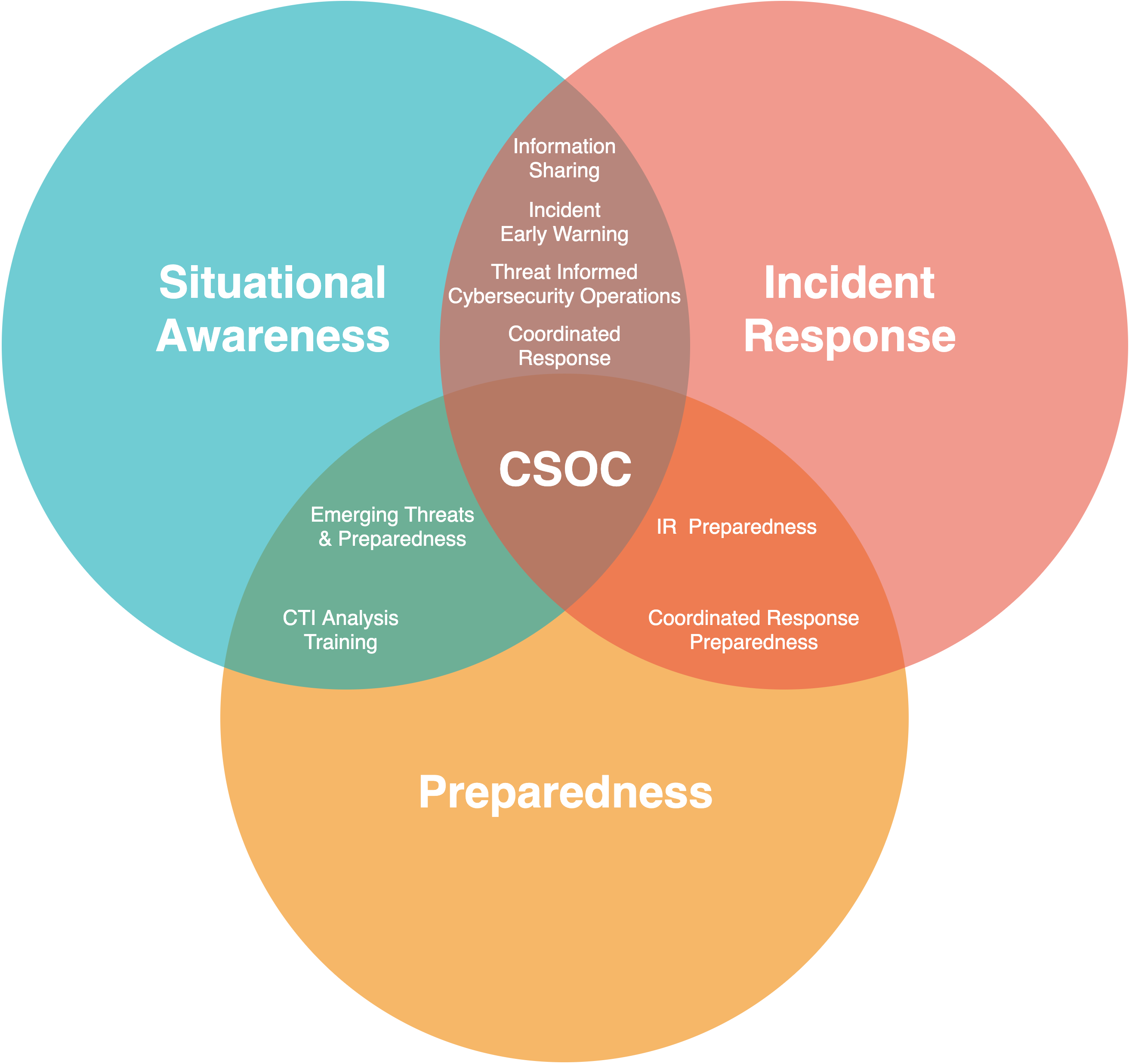}
	\caption{High-Level CSOC Blueprint: Capacities and their Interplay.}
  \label{fig:concept}
 \end{center}
\end{figure}

To fully exploit and maximise the impact of (i), (ii) and (iii), it is necessary to establish feedback loops allowing a continuous exchange of insights generated by the activities taking place within each pillar. Most importantly, in support of interoperability and seamlessly cross-cutting the pillars, we advise pursuing a standards-based implementation approach to facilitate the demanded capacity for \textbf{shared} situational awareness, \textbf{coordinated} incident response, and \textbf{joint} preparedness within and across national borders. The above key pillars are presented in Figure \ref{fig:concept} and are described in the following subsections.

\subsection{The Role of Interoperability and Standardisation in Collaborative CSOCs}
Open standards are specifications made available to the general public and are developed, approved, and maintained through a collaborative, consensus-driven process \cite{itu-standards-definition}. They aim to facilitate interoperability and data exchange among different products or services and are intended for widespread adoption. Open standards enable interoperability by design through standardised features, allowing implementers to focus solely on their product's purpose, properties, and quality.

In this context, implementing an EU-wide collaborative cybersecurity operations ecosystem requires adopting an open-standards-based interoperability strategy, including but not limited to standardised network protocols, data formats and exchange mechanisms, and semantically interoperable data and concept models that enable the free exchange of information, insights, analytics, and responses across CSOCs. In addition, an open-standards-based interoperability strategy mitigates the vendor/software lock-in nuisance that can be a major disruptor in implementing a flexible, sustainable, and interoperable CSOC infrastructure. For instance, from a shared threat situational awareness perspective, the EU CSOCs should leverage a standardised format for representing and exchanging cyber threat intelligence, information, and observables, designed to be agnostic of the system or system component operationalising the content. Similarly, defenders' tradecraft, standard operating procedures, playbooks and workflows should rely on and be encoded in an agreed-upon standardised format. The same principle applies to other operations and capacities, such as detection and logging, incident reporting and alerting, threat hunting, and command and control of defence systems. 

Further, CSOC services, tools, systems and system components, and processes that rely on open standards or may set a path towards standardisation must also comply with applicable EU policies \cite{certification} validating their level of security assurance.

It is worthy of note that open standards are a crucial element in boosting cybersecurity automation, allowing tools to interoperate natively and, simultaneously, minimising the need for customised integrations that are expensive and complex to build and maintain. Cybersecurity automation is a discipline that has been consistently identified as a key solution for enhancing the effectiveness and speed of security operations while reducing human errors.

\subsection{(Shared) Situational Awareness} \label{SA}
An essential capacity for protecting an organisation's assets against cyber threats is having accurate and up-to-date threat situational awareness, i.e., monitoring and understanding the relevant threat environment, assessing applicable risks to critical assets, and performing threat-informed decision-making for preparedness and anticipatory threat reduction. Considering also the potential commonalities and relations across intrusions, adversary tradecraft, and victimology, it becomes apparent that access to shared threat situational awareness is essential for the  Member States and the Union as a whole to achieve an enhanced security posture by leveraging the knowledge, experience, and capabilities of their allies. Shared threat situational awareness, as previously mentioned, is referred to as a critical cybersecurity capacity in the Cyber Blueprint to support the technical, operational, and strategic/political levels altogether.

Therefore, in this regard, CSOCs should incorporate an interoperable CTI toolbox comprising the necessary technical infrastructure to exchange, operationalise, generate and consume machine- and human-readable CTI, including open-source, commercial and from trusted groups, such as the reports provided by ENISA and the CSIRT network or, from a more international perspective, what is performed by the U.S. Cybersecurity and Infrastructure Security Agency (CISA) with the Automated Indicator Sharing capability \cite{AIS}. In addition, concerning relevant software, such as Threat Intelligence Platforms (TIPs), widely adopted but also open-source, include MISP \cite{MISP} and OpenCTI \cite{OpenCTI}; however, this work aims to remain software-agnostic and neutral and therefore does not recommend specific tooling. Instead, it recommends a standards-based approach to facilitate the aforementioned capacity and avoid software/vendor lock-in. For example, to the extent of our knowledge, the most comprehensive standard for representing and exchanging cyber threat information and intelligence in a machine-readable format is Structured Threat Information eXpression (STIX) \cite{stix2.1} and is currently operationalised by numerous CTI platforms and other tools. Further, for less technical consumers, standard document templating approaches (and associated style guides) should be introduced in support of interoperability, harmonisation and readability, and for optimally outlining, based on its type (e.g., sectorial threat landscape report), the key elements that should be contained to make it actionable and drive decision-making.

\begin{figure}[!h]
\begin{center}
	\includegraphics[width=\columnwidth]{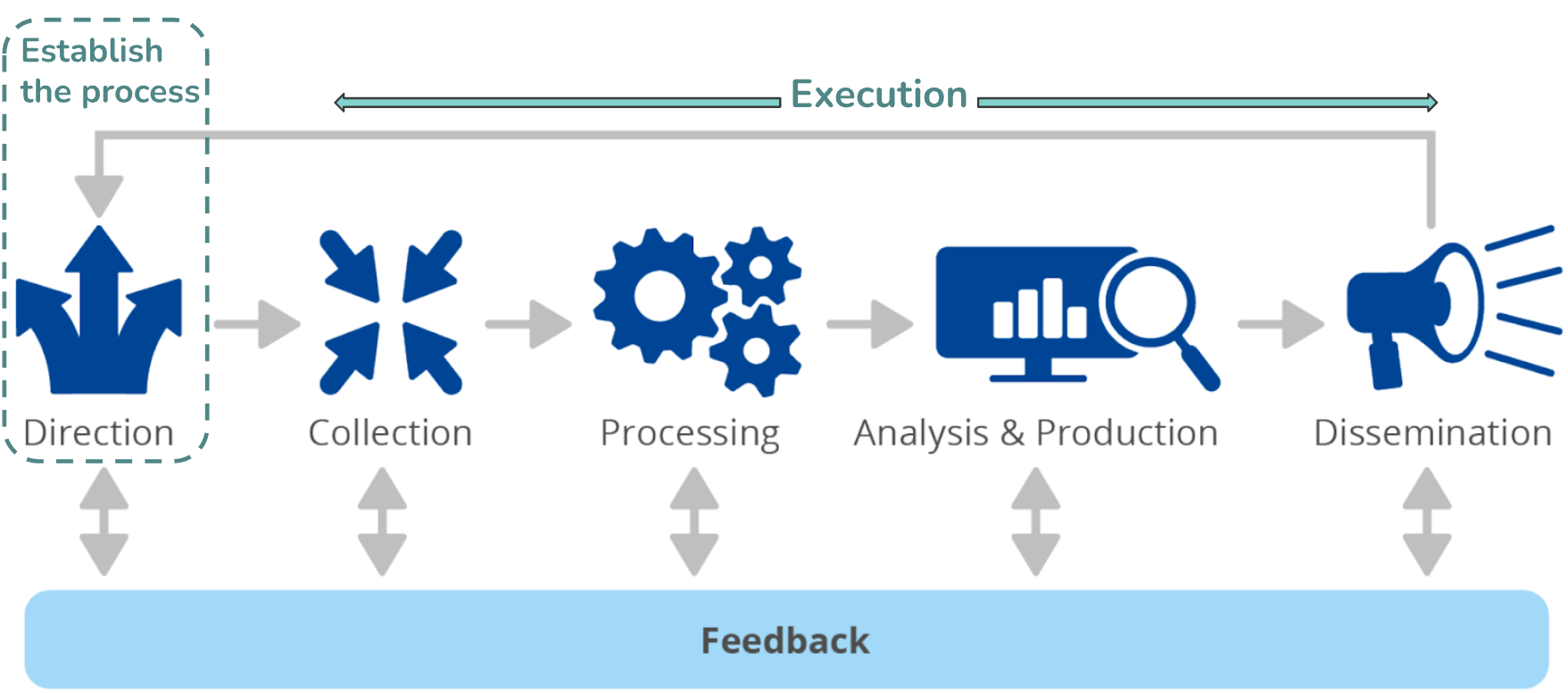}
	\caption{Intelligence Cycle \cite{ENISA-ETL} (slightly modified).}
 \label{fig:enisa-etl}
 \end{center}
\end{figure}

From a procedural perspective, practitioners adopt and adapt the intelligence cycle model, which provides a high-level framework in support of generating accurate, timely, relevant, and actionable intelligence. CSOCs that generate or consume CTI need to determine their stakeholders' intelligence requirements and build a tailored process that can address knowledge gaps. Concisely, the intelligence cycle comprises the following phases (see also Figure \ref{fig:enisa-etl}): 
\begin{itemize}
    \item Direction and planning that focuses on setting the overall process/program, including stakeholder management and the definition of intelligence and priority intelligence requirements.
    \item Collection of data, information, and other intelligence based on a defined collection management plan.
    \item Processing and exploitation of what has been collected to store everything in an easily accessible and processable manner.
    \item Intelligence analysis (structured analytical techniques are applied) to produce a product that answers an intelligence requirement. The type of the product is determined based on the nature of the stakeholder/consumer and its requirements.
    \item Dissemination of the product (intelligence) via predefined channels.
    \item Feedback mechanisms that provide the necessary information to refine the overall intelligence process.
\end{itemize}

\begin{figure*}[tbp]
\centering
	\includegraphics[width=0.9\hsize]{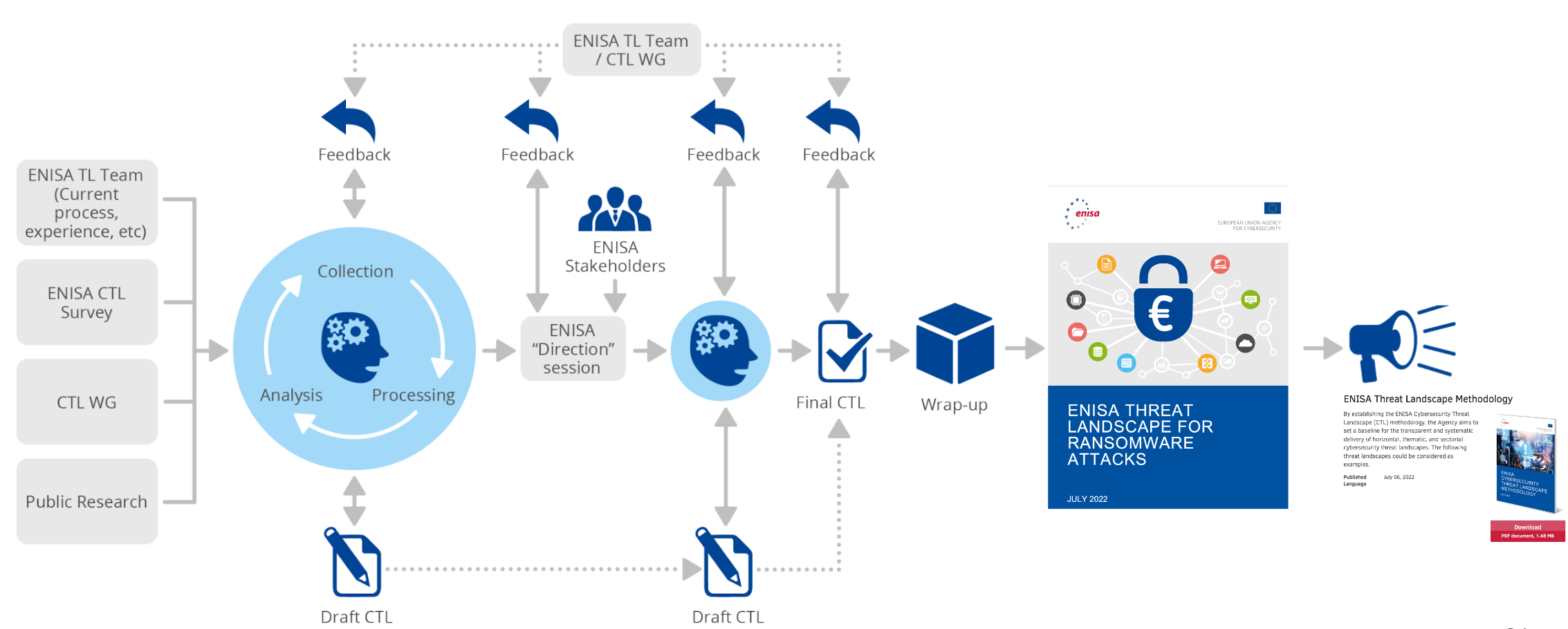}
	\caption{ENISA Cybersecurity Threat Landscape Methodology \cite{ENISA-ETL} (slightly modified).}
 \label{fig:enisa-etl1}
\end{figure*}

Figure \ref{fig:enisa-etl1} presents ENISA's adaptation of the intelligence cycle that drives the development of ETL reports.

\subsection{(Coordinated) Incident Response} \label{IR}
Effectively and efficiently performing incident response is a complex task and requires significant planning (technology, processes, people) and resources. A response can be preventative or triggered and can operationally intersect detection and analysis, containment, eradication, and recovery, including post-incident activities for improving the overall process \cite{NIST}. 

EU Member States are instructed to achieve a high common level of cybersecurity and, in particular, have the capacity and ability to respond effectively and efficiently to cyber incidents, including complying with reporting obligations. In addition, the EU Cyber Blueprint, in anticipation of incidents that can be classified as large-scale or incidents that cannot be contained solely by the affected Member State and consequently can affect the well-being of the Union, recommends having the capacity and ability to perform cross-border coordinated responses, meaning joint operations involving multiple Member States and associated EU actors, complemented by the activation of standardised technical, operational, and political modes of operation. In this context, CSOCs should incorporate incident response enablers that are in line with good practices established by the EU authorities, such as the CSIRTs Network and CERT-EU, while also providing the methods and processes needed to guarantee interaction with other CSOC instances, the relevant EU actors at the technical, operational, and political levels (e.g., CSIRTs Network, CERT-EU, ENISA, NIS Cooperation Group, CyCLONe \cite{CyCLONe}), and associated third parties (e.g., private entities). 

Interacting with the aforementioned established but diversified EU layers to enable coordinated response requires introducing standard operating procedures and technology that can support their execution and ensure compliance with policies and regulatory frameworks. Similarly, from a more technical perspective, incident responders can benefit from exchanging their tactics, techniques, and procedures, as well as other applicable knowledge and tooling (tradecraft), with peers nationally and internationally, enabling a more collective approach to cyber defence. The same approach can also mitigate asymmetries in the incident response capabilities of Member States and allow responders of lower operational maturity to execute well-defined and optimised incident response plans while also reducing the possibility of human error. For instance, a shared process of a Member State competent authority could describe a set of activities to detect, investigate, analyse, and contain a particular ransomware strain and reference a second standardised EU process that details the information-sharing method to be followed and reporting obligations while ensuring compliance with data privacy regulations. Similarly, a documented process that is aimed to be reusable, repeatable, and optimised could identify cooperation modes with other teams nationally and internationally who have the know-how to assist with particular activities (e.g., requests for information and support in specific functions such as malware reverse engineering for mobile applications). The efficacy of these coordinated and shareable incident response processes within the CSOC will be augmented by the enhanced shared situational awareness capacity pillar (see subsection \ref{SA}), as it will allow for the more prompt and accurate identification of causes, impact, origin and attribution of the incident (the latter being important for organisational but also strategic/political response measures). Work on implementing such mechanisms is ongoing \cite{mavroeidis2021integration, mavroeidis2022cybersecurity}.

To the extent of our knowledge, a relevant ongoing standardisation effort to support the aforementioned is the OASIS Collaborative Automated Course of Actions Operations (CACAO) \cite{cacao-spec} \cite{cacao-tc}. Briefly, CACAO defines a playbook schema and a taxonomy that standardise the way we create, document, and share defenders' cybersecurity operations processes and procedures. It is represented in a machine-readable format and is aimed to be vendor- and software-agnostic. CACAO can support the execution of standard operating procedures via the definition of executable orchestration workflows.


\subsection{(Joint) Preparedness} \label{Prep}
A key capacity-building activity in cybersecurity is the strengthening of defenders' capabilities by providing training on relevant topics (from designing security policies and security management to specialised training in tooling and procedural and methodological training pertinent to specific operational roles and responsibilities) tailored to users with different levels of expertise and maturity and also using a diversified set of delivery methods (from theoretical to hands-on training) that provide different learning opportunities. These training activities are also valuable in joint preparedness, facilitating capacity building regarding incident response processes, coordination mechanisms for responding to large-scale incidents, acquiring hands-on experience and transferring results and lessons learned from previous incidents and training sessions. In this context, the CSOC should integrate cybersecurity operations training features covering different training delivery methods tailored to the respective training goals and objectives. These could include:
\begin{enumerate}[label=(\roman*)]
\item \textit{online courses} delivered openly, for instance, through Massive Open Online Courses (MOOC) platforms or confidential/private channels for specific trusted groups such as the EU CSIRTs Network; 
\item \textit{table-top exercises} such as the Blueprint Operational Level Exercise (Blue OLEx) \cite{blueolex2022}, which aims to further enhance the EU operational (i.e., CyCLONe's) coordination in case of a large-scale cyber incident/crisis, or such as the ones provided by CISA \cite{CISA_Table};
\item hands-on \textit{Cyber Range training programmes}, delivered through platforms such as the KYPO cyber range \cite{KYPO} leveraged by the CONCORDIA competence network and THREAT-ARREST \cite{THREAT-ARREST} developed in the context of the corresponding Horizon 2020 project.
\end{enumerate}

The above methods can cover the training on CTI and Incident Response enablers specified in subsections \ref{SA} and \ref{IR} and should receive continuous feedback from these activities that will allow the repositioning and refinement of the training accordingly, ultimately providing a more threat-informed approach that can maximise the impact of the training and help achieve compliance with the requirements of the NIS Directive. Also, a significant focus should be given to training standard operating procedures for cooperation at the technical, operational and political-strategic levels as defined in the Cyber Blueprint regarding large-scale incidents and crises.

\subsection{Capacity Interplay}
As briefly highlighted in the above subsections and as visualised in Figure \ref{fig:concept}, a key aspect of the proposed CSOC blueprint is the interplay of the three capacity pillars. In other words, the required capacities that will drive the provision of specific tools, processes, and people need to function synergistically, creating a state-of-the-art cyber defence ecosystem. For instance, the exchange of information and feedback between the Situational Awareness and Incident Response pillars can enable:
\begin{itemize}
    \item Early warning of threats and incidents for prevention or anticipatory threat reduction through early detection and shareable courses of action that may allow for mitigation or remediation. Defenders should exchange observables, indicators of compromise and behaviour, and higher-level intelligence about relevant adversaries and their goals using standard agreed-upon formats to simplify the encoding/documentation, exchange and consumption of intelligence. Similarly, defenders may perform a timely defence gap assessment in light of this information.
    \item Timely information sharing via interoperable technologies and robust processes and procedures that can be potentially automated. For instance, an indicator of compromise can trigger an IR workflow that also orchestrates and automates the generation and sharing of CTI objects with the rest of the community.
    \item Coordinated response through the continuous exchange of information while cooperating with other teams on a technical know-how mode or through formally documented IR processes that integrate and orchestrate response functions amongst peers within and across organisational and geographical boundaries.
\end{itemize}

Equivalently, the interplay between Preparedness, Situational Awareness, and Incident Response pillars can enable:
\begin{itemize}
    \item Identification, specification, and delivery of training that is relevant to the current threat landscape; driven by shared (consumer) and generated (producer) CTI and also informed by the actual incidents handled by IR teams.
    \item Increased capacity to train staff on CTI (such as intelligence analysis, performing assessments for situational awareness, and relevant tooling), as well as training on IR (such as exercises on cross-border coordinated IR procedures, tooling, and processes adopted within the CSOC).
    \item Continuously improving the Situational Awareness and Incident Response capacity (technology, process efficacy) and defenders' capabilities (training) based on trainee performance, and overall results of the training sessions.
\end{itemize}
Therefore, these feedback loops enable the interplay between the different CSOC capacity pillars, allow for cross-enhancement, and serve as additional enablers that increase the efficiency and effectiveness of cybersecurity operations capabilities of both individuals and the CSOC as a whole.

\section{Concluding remarks \& Next steps} \label{section:conclusions}
Stakeholders across sectors (from technical to operational and political levels) are increasingly aware of the fact that, as our reliance on digital technologies and services increases, so does the impact cybersecurity incidents can have on our economies, our societies, and our democracies at large. Specifically, within the EU, this has led to an increasingly complex and constantly shifting regulatory landscape, which has motivated the expansion of security operations but also has complicated (and is expected to complicate further; e.g., see NIS2) compliance, especially for OES and entities entrusted with the cybersecurity of Member States.

Considering the above, this work proposed a capacity-focused blueprint to support architecting and establishing interoperable CSOCs that holistically cover and combine shared threat situational awareness, coordinated incident handling, and joint preparedness. Particular emphasis was given to cross-border, cross-organisational, and cross-functional collaboration and coordination with an increased focus on OES, organisations entrusted with the cybersecurity of Member States, and relevant EU cybersecurity entities and bodies. Furthermore, this work stressed the interplay between the different capacity pillars as their synergy could maximise their impact and pinpointed that the realisation of a sustainable and resilient EU-wide interoperable CSOC infrastructure should account for a standards-based interoperability strategy at the design phase. 

The approach presented herein will be validated in the context of two EU-funded research projects focusing on increasing the EU's capacity in this regard. The project JCOP ("Joint Cybersecurity Operations Platform"; funded from the Connecting Europe Facility programme under Grant Agreement No. INEA/CEF/ICT/A2020/2373266) leverages the CSOC blueprint for the creation of a Southeast Europe Coordinated Response Cluster encompassing the National Cybersecurity Authorities of Greece (Hellenic Ministry of Digital Governance - National Cyber Security Authority) and Cyprus (Cypriot Digital Security Authority), with the Norwegian National Security Authority as a cross-validator. The project PHOENi$^2$X ("A European Cyber Resilience Framework With Artificial Intelligence -Assisted Orchestration \& Automation For Business Continuity, Incident Response \& Information Exchange"; funded from the European Union's Horizon Europe Framework Programme, under Grant Agreement No. 101070586) leverages the CSOC blueprint for creating a Cyber Resilience Framework tailored to the needs of OES and will be validated by OES in the energy, transport and healthcare sectors in Greece, Spain, and Cyprus, respectively. 

Consequently, the solutions currently being designed and developed, and later validated in the context of the two European projects, follow the principles sketched by the CSOC capacity-focused blueprint presented herein. Tangible evidence of its applicability will be collected, along with valuable feedback and pointers for its further refinement.

\section*{Acknowledgment}
This work has received funding from the European Union's Connecting Europe Facility (CEF) programme under Grant Agreement No. INEA/CEF/ICT/A2020/2373266 (JCOP project) and the Horizon Europe programme under Grant Agreement No. 101070586 (PHOENi$^2$X project). Views and opinions expressed are, however, those of the authors only and do not necessarily reflect those of the European Union or the European Research Council Executive Agency. Neither the European Union nor the granting authority can be held responsible for them. In addition, this work was supported by CYENTIFIC AS (Organization No. 930 271 691).

\end{document}